\documentstyle[twoside,fleqn,espcrc2,,here,psfig]{article}

\newcommand{\be}{\begin{equation}}
\newcommand{\ee}{\end{equation}}
\newcommand{\bea}{\begin{eqnarray}}
\newcommand{\eea}{\end{eqnarray}}
\newcommand{\bean}{\begin{eqnarray*}}
\newcommand{\eean}{\end{eqnarray*}}

\newcommand{\ba}{\begin{array}}
\newcommand{\ea}{\end{array}}

\newcommand{\AmS}{{\protect\the\textfont2
  A\kern-.1667em\lower.5ex\hbox{M}\kern-.125emS}}

\title{The spin correlation in top quark production:\\
       QCD corrections vs anomalous couplings
      \thanks{Talk presented by J. Kodaira at 
        Loops and Legs 2000, Bastei, Germany, April 9-14, 2000.}}

\author{Yuichiro Kiyo, Jiro Kodaira, Kazushige Morii, Takashi Nasuno
       \address{Department of Physics, Hiroshima University, 
     Higashi-Hiroshima 739-8526, Japan}
         and
     Stephen Parke \address{Theoretical Physics Department,
          Fermi National Accelerator Laboratory, P. O. Box 500,
        Batavia, IL 60510, USA }}

\begin{document}

\begin{abstract}
We discuss top quark production and its subsequent decay
at lepton colliders including both QCD corrections and anomalous
$\gamma / Z - t \bar{t}$ couplings.
The off-diagonal spin basis for the top and anti-top quarks
is shown to be useful to probe the anomalous couplings.
\end{abstract}

\maketitle

\section{Introduction}
Since the discovery of the top quark, with a large mass \cite{TEVA},
its properties have been widely discussed to obtain a better
understanding of the electroweak symmetry breaking and to
search for hints of physics beyond the standard model (SM).
It has been known that top quarks decay before
hadronization~\cite{BIGI}.
Therefore there will be sizable angular correlations between
the decay products of the top quark and the spin of the
top quark~\cite{KUHN}.
Based on this observation, it is expected that
we can either test the SM or obtain some signal from new physics
by investigating the angular distributions of the decay products
from polarized top quarks.

Applying the narrow width approximation to the top quarks,
we can discuss the production process and decay process separately.
There are many papers on the spin correlations in top quark
production and also the angular distributions of decay products
by combining the production with the decay process~\cite{MP}.  
Although it was common to use the helicity basis to decompose
the top quark spin, it has been pointed out by Mahlon, Parke and
Shadmi~\cite{MAHL} that there is a more optimal decomposition
of the top quark spin depending on the process and the center of energy.

On the other hand, there are also many detailed studies on the
effects of new operators which might come from physics beyond
the SM~\cite{AnoCou}.
The fact that the SM is consistent with the data
within the present experimental accuracy tells us that
the size of the effects of new physics is at most comparable to
or smaller than the radiative corrections in the SM.
Therefore it might be important to estimate also the effects
of the SM radiative corrections.

In this talk, we will discuss the QCD corrections to the spin
correlations in the top quark productions at lepton colliders
and present the angular distribution of the decay product
including both the QCD corrections and the so-called anomalous
couplings for the $\gamma / Z - t \bar{t}$ interaction.

\section{QCD corrections to the spin correlation}
\begin{figure}[htbp]
\begin{center}
\leavevmode\psfig{file=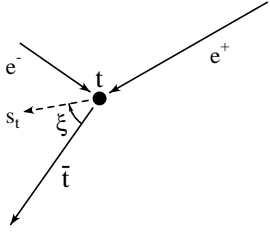,width=3.5cm}
\vspace{-0.6cm}
\caption{The spin vector ${\bf s}_t$ for the top quark is 
defined in the top quark rest frame.} 
\label{fig:spin}
\end{center}
\end{figure}

We first discuss the QCD correction to the spin dependent
cross section for the top quark production from
the polarized $e^- e^+$~\cite{KODA}. 
The spin direction of the top quark is parameterized by $\xi$
as in Fig.\ref{fig:spin} at the top quark rest frame.
The anti-top quark spin state is similarly defined by the same $\xi$.
%
%
The tree level cross section is easily calculated to be,
\[ \frac{d \sigma}{d \cos \theta^*}
        \left( e^-_L e^+_R \to t_{\uparrow} \bar{t}_{\uparrow} \right)
        = \frac{d \sigma}{d \cos \theta^*}
      \left( e^-_L e^+_R \to t_{\downarrow}
        \bar{t}_{\downarrow}\right)\]
\be
    = \left( \frac{3 \pi \alpha^2}{2 s} \beta \right)\,
         | ( A_{LR} \cos \xi - B_{LR} \sin \xi ) |^2 \label{tree1}
\ee
\[ \frac{d \sigma}{d \cos \theta^*}
        \left( e^-_L e^+_R \to t_{\uparrow} \bar{t}_{\downarrow} 
       \  {\rm or} \  t_{\downarrow} \bar{t}_{\uparrow} \right)\]
\[    = \left( \frac{3 \pi \alpha^2}{2 s} \beta \right)\,
         | ( A_{LR} \sin \xi + B_{LR} \cos \xi 
               \pm D_{LR} ) |^2 \]
where
\[  A_{LR} = \Bigl[ ( f_{LL} + f_{LR} ) \sqrt{1 - \beta^2}
            \sin \theta^* \Bigr] / 2 \]
\[  B_{LR} = \Bigl[ f_{LL} ( \cos \theta^* + \beta )
             + f_{LR} ( \cos \theta^* - \beta ) \Bigr] / 2 \]
\[  D_{LR} = \Bigl[ f_{LL} ( 1 + \beta \cos\theta^* )
             + f_{LR} ( 1 - \beta \cos\theta^* ) \Bigr] / 2 \]
\[  f_{IJ} = - Q_t + \frac{Q^e_I \, Q^t_J}{\sin^2 \theta_W}
              \frac{s}{s - M_Z^2}\]
Here $\alpha$ is the QED fine structure constant,
$\theta^* (\beta)$ is the scattering angle (the speed) of top quark
in the zero momentum frame
and $Q^e_I (Q^t_J )$ is the electron (top quark) coupling to the
Z boson.
$M_Z$ is the Z-boson mass and $\theta_W$ is the Weinberg angle.
From eq.(\ref{tree1}), one can see that the choice
$\tan \xi = A_{LR} / B_{LR}$ results in a large asymmetry.
This spin basis is called \lq\lq off-diagonal\rq\rq\  basis~\cite{MAHL}.
Since $|f_{LL}| \gg |f_{LR}|$ numerically, it turns out that only one spin
configuration dominates the cross section.
On the other hand, in the helicity basis ($\cos\xi = \pm 1$)
all configurations significantly contribute to the cross section.

The QCD corrections might modify the tree level results
since they induce an anomalous $\gamma / Z$ magnetic moment
for the top quark and allow for single real gluon emission.
Since the top and anti-top quarks are not necessarily produced
back to back at the one loop level, we discuss the single
spin correlation.
Here we show only the numerical results of our
calculations~\cite{KODA}.

Table I summarizes the strong coupling constant $\alpha_s$,
$\beta$,  and the 
tree and ${\cal O}(\alpha_s )$ level total
cross sections in $e^{-}_{L} e^{+}$ scattering.
\begin{center}
\begin{tabular}{|c||c|c|}
\hline 
$\sqrt{s}$    & $400$ GeV & $800$ GeV  \\ \hline
\hline
$\beta$                  & $0.484$  & $0.899$  \\ \hline
$\alpha_{s}(s)$          & $0.0980$ & $0.0910$  \\ \hline
$\sigma_{Total}$ Tree (pb)
                         & $0.8707$ & $0.3531$  \\ \hline
$\sigma_{Total}$ $\cal{O}\mit (\alpha_{s})$ (pb)
                         &  $1.113$ & $0.3734$  \\ \hline
\end{tabular}
\\[0.1in]
Table \ I: The values of $\beta$, $\alpha_s$, tree and next to leading order
           cross sections.
\end{center}
To see the effects of the QCD corrections to the spin correlation,
we write the cross section as a sum of two terms.
\bean
 \lefteqn{\frac{d \sigma}{d \cos \theta^*}
                  ( e_L^-  e_R^+ \to t_{\uparrow\downarrow} X )}\\ 
      &=& ( 1 + \kappa )\, \frac{d \sigma^0}{d \cos \theta^*}
                  ( e_L^-  e_R^+ \to t_{\uparrow\downarrow} X )\\
      & & \qquad\qquad  + \frac{d \sigma^R}{d \cos \theta^*}
                  ( e_L^-  e_R^+ \to t_{\uparrow\downarrow} X )
\eean
The first term is a part which is proportional to the tree level
cross section.
Therefore, $1 + \kappa$ is simply the multiplicative enhancement (K-) factor.
Whereas the second term give the ${\cal O}(\alpha_s )$
deviations to the spin correlations.
Our numerical studies show that
the ${\cal O} (\alpha_s )$ QCD corrections enhance the 
tree level result (the first term) and only slightly modifies the spin
orientation of the produced top quark (the second term).
The ratio of the second term to the first one is of order a few
percent.
In the kinematical region where the emitted gluon has small
energy, it is natural to expect that the real gluon emission effects
introduce only a multiplicative correction to the tree level result.
Therefore only \lq\lq hard\rq\rq\ gluon emission could possibly modify
the top quark spin orientation. 
What we have found, by explicit calculation, is that this effect
is numerically very small. 
In Table II, we give the fraction of the top quarks in the
subdominant spin configuration with $K$ factor for $e^-_L e^+$ scattering,
\[ \sigma 
\left(
     e^{-}_{L} e^{+} \rightarrow t_{\downarrow\uparrow} X(\bar{t}
 {\rm ~or~} \bar{t} g) 
\right) / \sigma^{Total}_{L} \]
for the helicity and off-diagonal bases.
These results suggest that the soft gluon approximation (SGA) will be
sufficient to estimate the 1-loop QCD corrections.
Actually, we have checked that the SGA can reproduce the full results
quite accurately by choosing an appropriate cut off $\omega_{\rm max}$
for the soft gluon.
The difference between the SGA using this $\omega_{\rm max}$ and the full
1-loop correction is smaller than the expected size of the 2-loop corrections.
\begin{center}
\begin{tabular}{|c||c|c|c|}\hline
   $\sqrt{s}$ ({\rm GeV}) &  $\kappa$  &
           Helicity  &  Off-Diagonal \\ \hline\hline
           &     &  \raisebox{-1.5ex}{0.336}  &  \raisebox{-1.5ex}{0.00124} \\
    400    &  0.278 &           &            \\ 
           &        &  \raisebox{1.5ex}{0.332} & 
               \raisebox{1.5ex}{0.00150} \\ \hline\hline
           &        &  \raisebox{-1.5ex}{0.168}  & \raisebox{-1.5ex}{0.0265} \\
    800    &  0.057 &           &            \\ 
           &        &  \raisebox{1.5ex}{0.165}
              & \raisebox{1.5ex}{0.0319}  \\ \hline
\end{tabular}
\\[0.1in]
Table \ II: The fraction of the $e^-_L e^+$ cross section in the
           subdominant spin. The upper (lower) line corresponds
           to the tree (one-loop) level.
\end{center}

\section{Decay distribution with anomalous coupling}
Although the QCD correction to the top quark production
is not so large, it should be included to detect \lq\lq small\rq\rq\ 
signals from possible new physics beyond the SM.
We analyze the top quark production and its subsequent decay
at lepton colliders including both QCD corrections and anomalous
$\gamma / Z - t \bar{t}$ couplings.

The process we are considering now is, in principle, a very complicated
$e^- e^+ \to 6$ one.
However, it has been known that the narrow width approximation
for the top quark, which is valid for $\Gamma_t \ll m_t$ 
(1.02 $\leq \Gamma_t \leq$ 1.56 GeV for 160 
$\leq m_t \leq$ 180 GeV), makes the situation very simple.
Namely, we can separate the physics into the top production
and the decay density matrices~\cite{JEZA}.

Let us first discuss the top quark production (density matrix).
We assume a general form for the $t$-$\bar{t}$-$Z/\gamma$ vertex as, 
\bea
  \Gamma^V_\mu &=&
   g^V \left\{ \gamma_{\mu} \left[ Q_L^V \omega_{-} + Q_R^V \omega_{+}
            \right]\right. \nonumber\\  
   &&  \qquad + \left.\frac{(t - {\bar{t}})_\mu}{2 m_t} 
     \left[ G_L^V \omega_{-} + G_R^V \omega_{+} \right] \right\}
                 \label{gvp}
\eea
where $t, \bar{t}$ are momenta of the top and anti-top quarks, 
$m_t$ is the top mass, $\omega_{-}/\omega_{+}$ is
the left/right projection operator,
and $V= Z$ or $\gamma$. 
For the $e$-$\bar{e}$-$Z/\gamma$ vertex, we use the well 
established SM interaction.
At the tree level in the SM, the coupling constants 
$G^{V}_{L,R}$ are zero.
The combination of form factors  
$G^{\gamma, Z}_R+G^{\gamma, Z}_L \equiv f_2^{\gamma, Z}$ is 
induced even at the one-loop level in the SM.  
Whereas, another combination $G^{\gamma, Z}_R - G^{\gamma, Z}_L
\equiv i f_3^{\gamma, Z}$ which is 
related to a CP violating interaction, called electric 
and weak dipole form factors (EDM and WDM) appears as, at least,
the two-loop order effect.
Thus they are negligibly small and non-zero value of $f_3^{\gamma,Z}$ 
is considered to be a contribution from new physics beyond the SM. 
We presume some non-zero value for $f_3^{\gamma,Z}$
and consider the top production and its decay.
The QCD one-loop correction is easily incorporated into this
analysis if one remembers that the one loop effect is very well
approximated by the SGA.
In the SGA, QCD effects can be expressed by the modified
$t$-$\bar{t}$-$Z/\gamma$ vertex~\cite{KODA}, eq.(\ref{gvp}).
\[  Q_{LR}^V \sim 1 + {\cal O}(\alpha_s ) \quad , \quad
  G_R^V + G_L^V \sim {\cal O}(\alpha_s )\]
\[ G_R^V - G_L^V = {\cal O}(\alpha_s^2 ) \sim 0 .\]
The top quark production amplitudes now read,
\bea
\lefteqn{M(e_L^-e_R^+\rightarrow t_\uparrow \bar{t}_\uparrow,
                        t_\downarrow \bar{t}_\downarrow)} \nonumber\\
&=&  \mp 4\pi \alpha
\left[ \hat{A}_{LR} \cos \xi - \hat{B}_{LR}
\sin \xi \pm i E_{LR} \right] \nonumber\\
\lefteqn{M(e_L^-e_R^+\rightarrow t_\uparrow \bar{t}_\downarrow, 
                        t_\downarrow \bar{t}_\uparrow)}\nonumber\\
&=& \ \ 4\pi \alpha
\left[ \hat{A}_{LR} \sin\xi +  \hat{B}_{LR}\cos\xi
 \pm \hat{D}_{LR} \right].  \nonumber\\
&&  \label{proamp}
\eea
where we have chosen the phases of spinors to be real.
The coefficients with \ $\hat{}$ \  receive the contribution from the QCD
corrections,
\[  \hat{C}_{LR} = C_{LR} + {\cal O}(\alpha_s ) .\]
For the explicit expressions, see ref.\cite {KODA}.
The function $E_{LR}$ linearly depends on $f_3^{\gamma,Z}$
and is given by,
\[ E_{LR} = \frac{1}{2} (h_{LL} - h_{LR}) \frac{\beta \sin \theta^* }
            {\sqrt{1-\beta^2}} \ ,\]
with
\[ h_{IJ} = -G^{\gamma}_J (t) + \frac{Q^Z_I (e) G^Z_J (t) }{\sin^2\theta_W} 
      \frac{s}{s - M_Z^2} \ .\]

The problem now is how to detect the anomalous coupling in
the top quark events.
It is easily understood that the effects of the anomalous coupling
on the top quark production cross sections should be small and
undetectable since the anomalous coupling is assumed to be comparable
to or smaller than the QCD correction in size and we already know
the QCD correction itself to be small.
Therefore we consider the angular distribution of top decay products
which depends linearly on $f_3^{\gamma,Z}$.

In the decay process, we assume V-A interaction of the SM in 
$t$-$b$-$W$ vertex. We employ the semi-leptonic decay, 
$t\rightarrow b W\rightarrow b \bar{l}\nu$ for simplicity. 
Neglecting the mass of the final state fermions,
the decay amplitude $D_{s_{t}}$ 
(for $t_{s_t} \to b \bar{l} \nu$) is known to be given by 
\bea
   D_{\uparrow} &=&
\frac{2 g^2 V_{tb} \sqrt{ b \cdot \nu ~m_t E_{\bar{l}}}
      }{2 \nu \cdot \bar{l} - M_W^2 + i M_W \Gamma_{W}} 
\cos\frac{\theta_{\bar{l}}
          }{2} \nonumber\\
D_{\downarrow} &=&
\frac{ 2 g^2 V_{tb} \sqrt{ b \cdot \nu ~m_t E_{\bar{l}}}
      }{ 2 \nu \cdot \bar{l} - M_W^2 + i M_W \Gamma_{W} } 
\sin\frac{ \theta_{\bar{l}}
         }{2} e^{ -i \phi_{\bar{l}} } \nonumber\\
& & \label{eq:D2}
\eea 
where the names of final particles are used as substitute
for their momenta.
$M_W$ and $V_{tb}$ are the masses of the
W boson and the Cabbibo--Kobayashi--Maskawa (CKM) matrix.

\vspace{-0.8cm}
\begin{figure}[htbp]
\begin{center}
\leavevmode\psfig{file=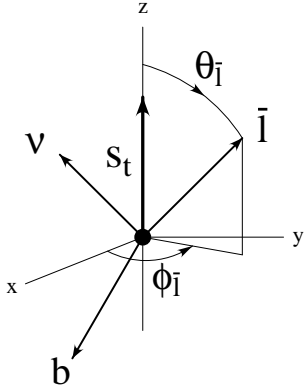,width=4cm}
\vspace{-0.6cm}
\caption{The definition of polar and azimuthal angles.}
\label{fig:angle}
\end{center}
\end{figure}

\vspace{-0.8cm}
\noindent
The polar and azimuthal angles of the $\bar{l}$ momentum
($\theta_{\bar{l}}, \phi_{\bar{l}}$) are defined 
in the top quark rest frame, 
in which $z$-axis coincides with the chosen spin axis $s_t$
and $x-z$ is the production plane, Fig.\ref{fig:angle}.
We have a similar expression $\bar{D}_{\uparrow\downarrow}$
also for the anti-top quark decay.

Now, the differential cross-section for the process $e^- e^+ \to 
t\bar{t}$ followed by the decays $t \to X_t\,,\,\bar{t} \to \bar{X}_t$
is described in terms of the production
and decay density matrices
$\rho_{s_t \bar{s}_t , s'_t \bar{s}'_t}$\, ,\,
$\tau_{s_t s'_t}$ and $\bar{\tau}_{\bar{s}_t \bar{s}'_t}$ as,
\bean
  \lefteqn{d \sigma \left( e^- e^+ \to t\bar{t} 
               \to X_t \bar{X}_t \right)}\\
     &\propto& \sum_{s_t, \bar{s}_t,s'_t,\bar{s}'_t}
           \rho_{s_t \bar{s}_t , s'_t \bar{s}'_t}
          \tau_{s_t s'_t} \bar{\tau}_{\bar{s}_t \bar{s}'_t} d L ,
\eean
where $d L$ is the phase space of the final particles
and the density matrices can be obtained from
eqs.(\ref{proamp}) and (\ref{eq:D2})~\cite{JEZA}. 
\[ \rho_{s_t \bar{s}_t , s'_t \bar{s}'_t}
          = M_{ s_t \bar{s}_t} M^{\ast}_{s'_t \bar{s}'_t} \]
\[  \tau_{s_t s'_t} \propto 
     D_{s_t} D^\ast_{s'_t} \propto   \left( \ba{cc}
        1+\cos\theta_{\bar{l}} & \sin\theta_{\bar{l}} e^{i\phi_{\bar{l}}}\\
       \sin\theta_{\bar{l}}e^{-i\phi_{l}}& 1-\cos\theta_{\bar{l}} \\
    \ea \right)_{s_t s'_t} \]
$\bar{\tau}_{\bar{s}_t \bar{s}'_t}$ is also given by the similar
expression.
When we calculated the production density matrix,
we have kept terms up to linear in $\alpha_s$ and $f_3^{\gamma,Z}$
for the consistency of our approximation.
We have also applied the narrow width approximation
for the W boson in eq.(\ref{eq:D2}) to derive the above result.
From this expression, we see that there are terms which linearly 
depend on $f_3^{\gamma, Z}$ in the angular distributions of the 
charged leptons and the interference terms between amplitudes for 
different spin configuration play an important role.

\begin{figure}[H]
\begin{center}
\begin{tabular}{cc}
\leavevmode\psfig{file=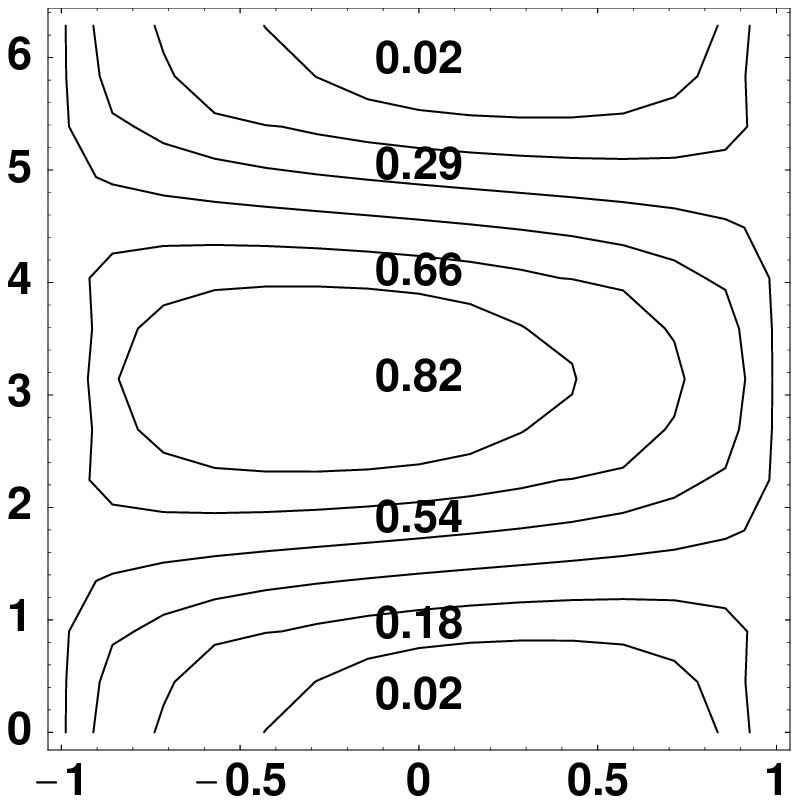,width=3.6cm} \hspace*{-0.4cm} & \hspace*{-0.4cm}
\leavevmode\psfig{file=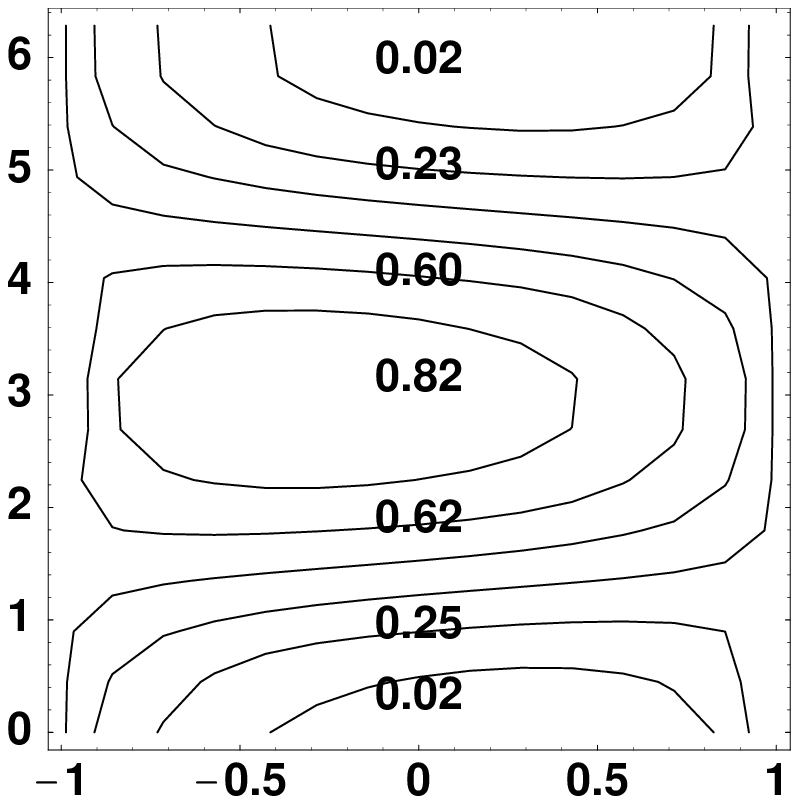,width=3.6cm}
\end{tabular}
\vspace{-0.6cm}
\caption{The double differential cross section
$d^2 \sigma / d \cos \theta_{\bar{l}} d \phi_{\bar{l}}$
in the helicity basis.
The left (right) figure correspond to the
cross section without (with) the anomalous $f_3^{\gamma,Z}$
coupling. Vertical and horizontal axes
correspond to the azimuthal $\phi_{\bar{l}}$ and 
the polar angle $\cos\theta_{\bar{l}}$, respectively.}
\label{helcorr}
\end{center}
\end{figure}

\begin{figure}[H]
\begin{center}
\begin{tabular}{cc}
\leavevmode\psfig{file=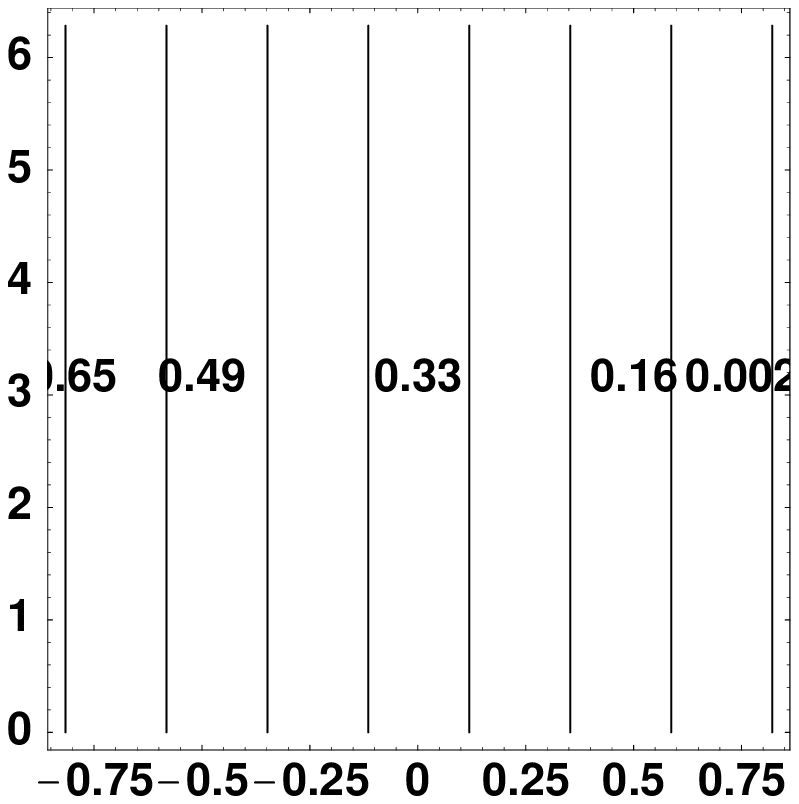,width=3.6cm} \hspace*{-0.4cm} & \hspace*{-0.4cm}
\leavevmode\psfig{file=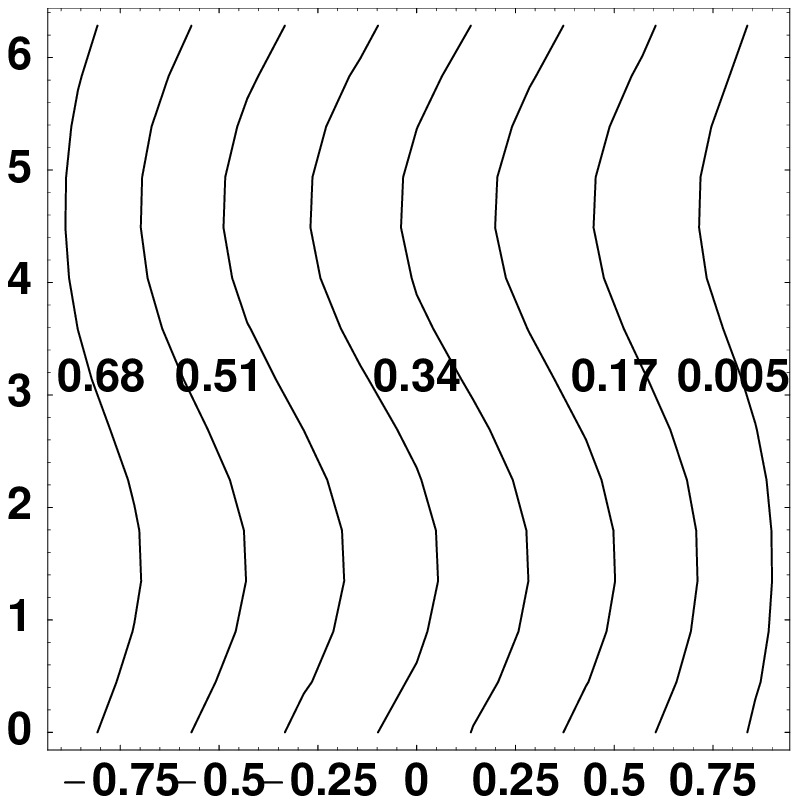,width=3.6cm}
\end{tabular}
\vspace{-0.6cm}
\caption{The double differential cross section
in the off-diagonal basis.
The left (right) figure correspond to the
cross section without (with) the anomalous $f_3^{\gamma,Z}$
coupling.
The axes are the same as in Fig.\ref{helcorr} }
\label{ODBcorr}
\end{center}
\end{figure}

Here we take an advantage of the freedom for the choice of the 
spin basis to detect the effect of the anomalous couplings.
Note that the differential cross section itself is (should be)
independent of the choice of the spin basis.
However, the \lq\lq choice of the variables\rq\rq\  can depend on
the spin basis.
We have calculated the angular distribution of $\bar{l}$
in the top quark decay after integrating out other variables.
We plot the $\theta_{\bar{l}} - \phi_{\bar{l}}$ correlations
both in the helicity (Fig.\ref{helcorr})
and the off-diagonal basis (Fig.\ref{ODBcorr}).
We set $\sqrt{s} = 400 \ GeV$ and assume $f_3^{\gamma,Z}= 0.2$
just for an illustration.
The both figures are for $\cos \theta^* = 0$.
However the pattern of the correlation is essentially the same for
all scattering angles.
One can see that it is very hard to identify the effects of the
anomalous couplings in Fig.~\ref{helcorr}, 
This situation changes drastically if we take the
off-diagonal basis (Fig.~\ref{ODBcorr}).  
As the SM result produces almost no 
azimuthal angular dependence in this 
basis (these azimuthal angular dependencies are caused by interferences 
effects in a given spin basis and these are very small in the 
off-diagonal basis), 
we recognize the effect 
of the anomalous coupling as a deviation from the flat distribution.
For the value of the anomalous coupling 
we have chosen, these new interactions change the shape nearly by 10\%.

\vspace{-0.8cm}

\vspace{-2cm}

\begin{figure}[H]
\begin{center}
\leavevmode\psfig{file=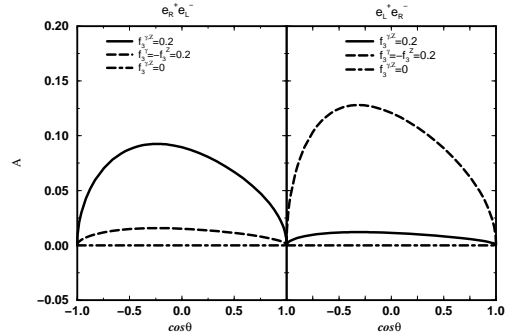,width=7.5cm,angle=-90}
\vspace{-2cm}
\caption{Azimuthal asymmetry as a function of $\cos \theta^*$ 
in the off-diagonal basis.} 
\label{fig5} 
\end{center}
\end{figure}

\vspace{-0.8cm}
In order to show the effect of the $f_3^{\gamma,Z}$ 
more clearly, we partially integrate the cross section over the 
azimuthal angle and define the azimuthal asymmetry. 
Let $\sigma^{1,2}$ denote the partially integrated cross-sections 
over the azimuthal angle,
%
\bean
 \sigma^1 (\theta^*) &=& \int_{0}^{\pi} d \phi_{l} 
       \left( \frac{d \sigma}{d \cos\theta^* d \phi_{l}} \right)\ ,\\
 \sigma^{2} (\theta^*) & =& \int_{\pi}^{2 \pi} d \phi_{\bar{l}} 
       \left( \frac{d \sigma}{d \cos\theta^* d \phi_{l}}\right)
\eean
where other variables have been integrated out already.
We define the azimuthal asymmetry in order to pull out the 
effect of anomalous interactions,
\[   {\cal A} (\theta^*)  = 
  \frac{\sigma^{2} (\theta^*) -\sigma^{1} (\theta^*)}
    {\sigma^{2} (\theta^*) + \sigma^{1}(\theta^*)} . \]
We plot the asymmetry as a function of $\cos\theta^*$ in 
Fig.\ref{fig5} at 
$\sqrt{s} = 400 GeV$ for the $e^+_R e^-_L$ and $e^+_L e^-_R$ 
annihilation.  

\vspace{-0.8cm}

\begin{figure}[H]
\begin{center}
\leavevmode\psfig{file=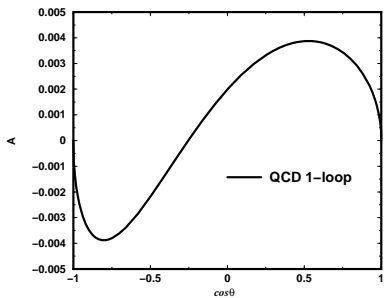,width=5.5cm}
\vspace{-0.6cm}
\caption{Azimuthal asymmetry induced by the QCD correction
in the off-diagonal basis.} 
\label{QCDasym} 
\end{center}
\end{figure}

\vspace{-0.8cm}
\noindent
In this figure, the dot-dashed line comes from 
the SM (with QCD corrections) and others from anomalous couplings.
At the SM tree level, the asymmetry is exactly zero and
the QCD radiative corrections induce a numerically negligible
asymmetry as shown in Fig.\ref{QCDasym}.
The asymmetry strongly depends on the value and the sign 
of $f_3^{\gamma,Z}$.
In the case of $e^+_R e^-_L$, the effects 
of the anomalous interactions $f_3^\gamma $ and $f_3^Z$ 
are additive and have a larger asymmetry 
when their signs are the same.
But when their signs are opposite, 
these effects become subtractive and lead to a smaller asymmetry.  
This feature changes in the case of $e^+_L e^-_R$.  
In the off-diagonal basis, the anomalous couplings 
produce the asymmetry of the order 10\%.
In the helicity basis, however, the deviation from the SM 
is only around 1.5\% since there exists some amount of asymmetry 
already in the SM. 

\vspace{-0.8cm}

\vspace{-0.8cm}

\section{Conclusion}
We have studied the top quark pair production and subsequent decays 
at lepton colliders.
First, we reported that the contribution of QCD corrections
is mainly just the enhancement of the tree level result
(K-factor) and does not change the spin configuration
of produced top quarks.
For a realistic next lepton colliders, let us say 
$\beta \sim 0.5$, the helicity basis is a poor choice
since all spin configurations contribute to the
production process.  
This means that there is a significant interference
between the intermediate spin states.
On the other hand, the off-diagonal basis is
a good choice since the contribution from
some spin states is zero or negligible even after
including the QCD corrections. 
This small interference makes the correlations between decay products
and the top spin very strong.
Using this advantage, we analyzed, secondly, the angular dependence
of the decay product of the top quark including
both the QCD corrections and the anomalous couplings. 
We have shown that the asymmetry amount to the order of 10\%
in the off-diagonal basis with chosen parameters which
may be detectable.

Although we have considered the anomalous couplings only
for the production process and showed the results
for their particular values,
the inclusion of new effects to the decay process
and more detailed phenomenological analyses for various
choices of the new interactions
are quite straightforward exercises.


\end{document}